# Li$^+$ solvation in pure, binary and ternary mixtures of organic carbonate electrolytes.

*Ioannis Skarmoutsos*[†,*], *Veerapandian Ponnuchamy*[†, ‡], *Valentina Vetere* [‡], *Stefano Mossa*[†,*]

[†] *Univ. Grenoble Alpes, INAC-SPRAM, F-38000 Grenoble, France; CNRS, INAC-SPRAM, F-38000 Grenoble, France and CEA, INAC-SPRAM, F-3800 Grenoble, France Emails: ioannis.skarmoutsos@cea.fr , stefano.mossa@cea.fr*

[‡] *Univ. Grenoble Alpes, LITEN-DEHT, F-38000 Grenoble, France and CEA, LITEN-DEHT, F-38000 Grenoble, France*

## Abstract

Classical molecular dynamics (MD) simulations and quantum chemical density functional theory (DFT) calculations have been employed in the present study to investigate the solvation of lithium cations in pure organic carbonate solvents (ethylene carbonate (EC), propylene carbonate (PC) and dimethyl carbonate (DMC)) and their binary (EC-DMC, 1:1 molar composition) and ternary (EC-DMC-PC, 1:1:3 molar composition) mixtures. The results obtained by both methods indicate that the formation of complexes with four solvent molecules around Li$^+$, exhibiting a strong local tetrahedral order, is the most favorable. However, the molecular dynamics simulations have revealed the existence of significant structural heterogeneities, extending up to a length scale which is more than five times the size of the first coordination shell radius. Due to these significant structural fluctuations in the bulk liquid phases, the use of larger size clusters in DFT calculations has been suggested. Contrary to the findings of the DFT calculations on small isolated clusters, the MD simulations have predicted a preference of Li$^+$ to interact with DMC molecules within its first solvation shell and not with the highly polar EC and PC ones, in the binary and ternary mixtures. This behavior has been attributed to the local tetrahedral packing of the solvent molecules in the first solvation shell of Li$^+$, which causes a cancellation of the individual molecular dipole vectors, and this effect seems to be more important in the cases where molecules of the same type are present. Due to these cancellation effects, the total dipole in the first solvation shell of Li$^+$ increases when the local mole fraction of DMC is high.

### 1. Introduction

In recent years the demand for portable power applications has been highly increased, thus giving a considerable impetus to the development of novel electrochemical devices, such as electric double layer capacitors and lithium ion batteries. Lithium ion secondary batteries are very common in consumer electronics, such as laptop computers and cell phones, while they are also growing in popularity for automotive applications in order to decrease the green-house gas emissions in the atmosphere and, hence, to prevent global warming[1-3]. In general Li-ion batteries have been deployed so far in a wide-range of energy storage applications, ranging from energy-type batteries of a few kilowatt-hour in residential systems with rooftop photovoltaic arrays to multi-megawatt containerized ones, for the provision of grid ancillary services.

Li-ion cells mainly employ intercalation materials as positive and negative electrodes and aprotic electrolytes to conduct $Li^+$. The chemical nature of the electrodes determines the energy output, while the electrolyte affects the rate of the energy release by controlling mass transport properties within the battery[1]. The interactions between the electrolyte and the electrode materials are also very important and the formation of electrified interfaces between them often dictates the performance of the device. An electrolyte should meet a list of minimal requirements in order to be used in such devices. In general it should be a good ionic conductor and electronic insulator, have a wide electrochemical window, exhibit electrochemical, mechanical and thermal stability, be environmental friendly and inert to other cell components such as cell separators, electrode substrates, and cell packaging materials.

As mentioned above, the transport of $Li^+$ ions inside the electrolyte determines the rate of the energy transfer, which has been stored on the electrodes[4]. According to the literature the transport of $Li^+$ ions is controlled by a two-step mechanism involving the solvation of the ions by the solvent molecules, followed by the migration of the solvated ions[5]. A deeper understanding of the solvation of $Li^+$ ions may therefore act as a springboard towards the rational design of novel electrolytes with improved $Li^+$ conductivity.

By now, the most commonly employed strategy towards the rational design of electrolytes with optimal properties for battery applications is to use mixtures of cyclic and non-cyclic organic carbonates[6]. In such a way the high dielectric constant of cyclic carbonates is combined with the low viscosity of acyclic carbonates ensuring good performances under low temperature environments. On the other hand, the higher thermal stability of cyclic carbonates ensures a reasonable operating temperature range for the mixed solvent.

Although several experimental and theoretical studies devoted to the interactions of $Li^+$ with pure and mixed carbonate-based electrolytes have already been published, the solvation structure and dynamics of lithium cations in these solvents is still a subject of debate. Interestingly, even the determination of the coordination number around the lithium ions in pure carbonate-based solvents has not been definitely resolved[7-18]. While the generally accepted picture comprises a tetrahedral coordination of the carbonyl oxygen atoms around $Li^+$, some experimental and theoretical studies propose the existence of local structures exhibiting slightly higher coordination numbers. The dependence of this local coordination number on the ion concentration is also somehow controversial. However, it should be emphasized that designing experimental methods or theoretical models to analyze the experimental data in order to provide a direct measurement of the coordination number is an extremely complicated task[7,19,20]. On the other hand, since the validation of molecular simulation results strongly depends on the direct comparison with experimental data, the development of experimental methods proving a direct determination of the coordination number becomes indispensable in order to obtain a clear picture about the local structural effects in liquid solvents.

On the basis of the above considerations the aim of the present study is not to give a final answer to this particular problem. The main purpose is to provide some general insight concerning the differences in the solvation mechanisms of Li$^+$ in pure and mixed binary and ternary carbonate-based solvents, by employing molecular dynamics simulations and quantum chemical calculations. Also, recent advances in the investigation of the local solvation structure by means of both experimental and theoretical techniques will be discussed and compared with the findings of the present study. Such a discussion might be used as a springboard towards a better understanding of the solvation phenomena in these electrolytes, significantly improving the rational design of electrolytes for battery applications.

## 2. Computational Methods and Details

### 2.1 Density Functional quantum chemical calculations

Quantum chemical calculations for several clusters of pure, binary and ternary electrolytes including a lithium cation have been performed using the ADF software[21]. The Density Functional Theory (DFT) has been employed for the optimization of structures, using the PBE GGA[22] functional and a TZP (core double zeta, valence triple zeta, polarized) slater type orbital (STO) basis set. Our calculations for some representative clusters have revealed that the basis set superposition error (BSSE) corrections[23] are negligibly small. A similar observation has also been pointed out in recent DFT studies[24,25]. Frequency analysis has been carried out for each structure ensuring the absence of imaginary modes, and confirming each structure as a minimum on the potential energy surface. Zero-point energy (ZPE) corrections have been also taken into account. Thermodynamic quantities such as the entropy, enthalpy and free energy of the investigated clusters have been estimated at the temperature of 298.15 K.

### 2.2 Molecular Dynamics Simulations

In the present study the solvation structure of Li$^+$ at infinite dilution in pure and mixed carbonate-based solvents has been investigated via classical molecular dynamics (MD) simulations. The selected pure solvents were ethylene carbonate (EC), propylene carbonate (PC) and dimethyl carbonate (DMC). The binary EC-DMC mixture with a molar composition 1 EC : 1 DMC and the ternary EC-DMC-PC one with 1 EC : 1 DMC : 3 PC molar composition were also studied.

Inside each simulation cubic box, one lithium cation was placed among 215 solvent molecules in the case of the pure solvents and the ternary mixture, whereas in the case of the binary mixture it was placed among 214 solvent molecules. Trial runs with larger system sizes have verified that the employed number of molecules are sufficient in describing the solvation structure of Li$^+$, avoiding in this way any possible artifacts arising from finite size effects. The initial configurations of the simulated systems were prepared by using the *Packmol* software[26].

The force fields employed in the simulations were adopted from previous studies[27,28]. The intermolecular interactions in these models are represented as pair wise additive with site-site

12-6 Lennard-Jones plus Coulomb interactions. The EC and PC molecules were kept rigid during the simulations (intramolecular geometries can be found in the Supporting Information). It should be mentioned that when investigating the solvation of $Li^+$ in pure solvents, test simulations were also performed using a flexible potential model[28] for EC. It was observed that the local structure and orientation around $Li^+$ in pure EC was not significantly affected compared to the case of the rigid model. This is possibly due to the fact that the rigid framework of cyclic carbonates is not significantly distorted by the presence of the intramolecular vibrational motions and therefore does not affect the packing of these molecules inside the solvation shell of $Li^+$.

In the case of DMC, due to the existence of different conformers in the gas and liquid phase[29-31], an intramolecular force field has been also employed. The intramolecular interactions have been represented in terms of harmonic angle bending and cosine series for dihedral angle internal rotations, whereas the bond lengths have been kept rigid by employing a modified version of the SHAKE[32,33] algorithm. The parameters of the intramolecular force field used are also presented in the Supporting Information. Due to the absence of a rigid framework, which is present in cyclic carbonates, intramolecular torsional motions could play an important role in the packing of DMC molecules. It should be noted however that conformational transitions were not detected on the timescale of our simulations.

As pointed out in the introduction, the generally accepted picture comprises a tetrahedral coordination of the carbonyl oxygen atoms around $Li^+$. The potential models used in the presented study predict such local structures[27,28] and for this reason they were selected to be employed in the simulations. Results obtained with different potential models already published in the literature will be also discussed together with the findings of the present study.

The equations of motion were integrated using a leapfrog-type Verlet algorithm with an integration time step of 1 fs[34]. The temperature has been fixed to 303.15 K and the pressure to 1 atm by coupling the systems to a Nose-Hoover thermostat and barostat with relaxation times of 0.2 and 0.5 ps, respectively[5,36]. The rigid body equations of motion for EC and PC molecules were expressed in the quaternion formalism[34]. A cut-off radius of 9.0 Å has been applied for all Lennard-Jones interactions and long-range corrections have also been taken into account. To account for the long-range electrostatic interactions the standard Ewald summation technique has been used[34]. The simulation runs were performed using the DL_POLY simulation code[37]. The systems were equilibrated for 25 ns and a subsequent run of 15 ns was performed in each case, in order to calculate equilibrium properties.

## 3. Results and Discussion

### 3.1 Relative stability of isolated clusters

By employing DFT calculations, using the methodologies described in section 2.1, a wide range of clusters of the type $Li(S)_n^+$ (where S=EC,PC,DMC and n=1-5) were optimized. Thermodynamic parameters, such as the enthalpy, entropy and binding energy of these clusters have been also estimated [38,39], based upon the calculation of the total partition function resulting from the contribution of the translational, rotational and vibrational degrees of freedom[39]. The binding energies of these clusters have been expressed in terms of the equation:

$$BE = E(Li(S)_n^+) - E(Li^+) - n\,E(S) \quad (1)$$

In order to provide an estimation of the preferable coordination number around the lithium cation, the most favorable path between these clusters $Li(S)_n^+$ has to be determined in terms of free energy changes. To do so, the free energies of the aggregates $Li(S)_n^+ + m\,S$ ( n+m=constant=5 in this case) were estimated, in order to define the aggregate exhibiting the lowest free energy value. In this way, if the fragment $Li(S)_n^+ + m\,S$ has the lowest free energy value, all paths of the type:

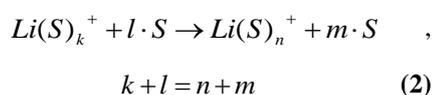

$$Li(S)_k^+ + l\cdot S \rightarrow Li(S)_n^+ + m\cdot S \,,$$
$$k+l = n+m \quad (2)$$

will exhibit a negative free energy change ΔG, thus indicating that all possible paths leading to the cluster $Li(S)_n^+$ are favorable and, therefore, this cluster could be considered the most preferable one in terms of the free energy changes. Note that the paths depicted in Eq. 2 can correspond to both additions or subtractions of solvent molecules from one particular cluster leading to another, depending on the relative difference $m-l$. If $m-l<0$, then the transition is being achieved through solvent addition, otherwise through solvent subtraction.

All the calculated values of enthalpy, entropy and binding energy of the investigated $Li(S)_n^+$ clusters, together with the free energy values of the $Li(S)_n^+ + m\,S$ aggregates are presented in Tables 1,2. From the results obtained it can be clearly seen that in the cases of EC and DMC, the predicted most favorable structures correspond to a tetracoordinated lithium cation. Similar conclusions have been drawn in the DFT study of Bhatt et al, studying the solvation of $Li^+$ in EC[24]. In the case of PC the free energy difference for the path going from $Li(PC)_3^+$ to $Li(PC)_4^+$ is very small. This is an indication that although the $Li(PC)_3^+$ cluster has been predicted to be the most favorable one, both structures could possibly exist in the bulk phase. The optimized structures for $Li(EC)_4^+$, $Li(PC)_3^+$ and $Li(PC)_4^+$, as well as for $Li(DMC)_4^+$ are shown in Figures 1-3. From the data presented in Table 1 it might be also seen that in the cases of the tetracoordinated clusters the binding energies decrease in the order $E(Li(PC)_4^+) > E(Li(EC)_4^+) > E(Li(DMC)_4^+)$, which is reasonable taking into account that the dipole moments of PC, EC and DMC molecules decrease in the same order (5.6, 5.3 and 0.35 D, respectively). It should be also noted that the entropic contributions to the free energy of each cluster exhibit the opposite trend, being more important in the case of the $Li(DMC)_4^+$.

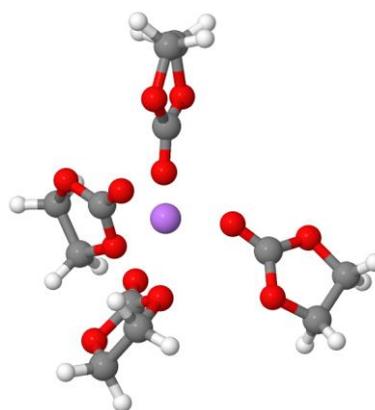

**Figure 1:** Structure of the optimized $Li(EC)_4^+$ cluster.

**Table 1:** Thermodynamical parameters and Binding energies of Li+ - Solvents clusters (H and BE are in kcal/mol and S in cal mol$^{-1}$ K$^{-1}$)

| Clusters | H | BE | S |
|---|---|---|---|
| EC | -1369.3 | | 72.8 |
| Li$^+$(EC) | -1293.6 | -47.6 | 83.2 |
| Li$^+$(EC)$_2$ | -2698.9 | -83.1 | 116.9 |
| Li$^+$(EC)$_3$ | -4090.4 | -104.6 | 142.2 |
| Li$^+$(EC)$_4$ | -5473.4 | -118.2 | 174.0 |
| Li$^+$(EC)$_5$ | -6849.9 | -126.9 | 214.6 |
| DMC | -1530.2 | | 82.5 |
| Li$^+$(DMC) | -1449.0 | -42.1 | 90.6 |
| Li$^+$(DMC)$_2$ | -3012.2 | -74.0 | 124.5 |
| Li$^+$(DMC)$_3$ | -4562.4 | -92.9 | 158.1 |
| Li$^+$(DMC)$_4$ | -6100.8 | -103.2 | 216.8 |
| Li$^+$(DMC)$_5$ | -7636.7 | -108.5 | 251.3 |
| PC | -1736.0 | | 79.6 |
| Li$^+$(PC) | -1662.2 | -49.5 | 89.8 |
| Li$^+$(PC)$_2$ | -3435.1 | -85.3 | 123.7 |
| Li$^+$(PC)$_3$ | -5193.2 | -106.5 | 154.5 |
| Li$^+$(PC)$_4$ | -6945.0 | -121.3 | 180.3 |
| Li$^+$(PC)$_5$ | -8687.2 | -128.8 | 231.3 |

H=Enthalpy, BE = Binding Energy, S = Entropy

**Table 2:** Gibbs free energy of clusters (kcal/mol)

| Clusters | G | | |
|---|---|---|---|
| | S=EC | S=DMC | S=PC |
| Li$^+$(S)+ 4S | -6882.4 | -7695.0 | -8728.1 |
| Li$^+$(S)$_2$ + 3S | -6906.7 | -7713.5 | -8751.3 |
| Li$^+$(S)$_3$ + 2S | -6914.8 | -7719.0 | -8758.8 |
| Li$^+$(S)$_4$ + S | -6916.2 | -7720.2 | -8758.5 |
| Li$^+$(S)$_5$ | -6913.9 | -7711.6 | -8756.2 |

Considering that the tetracoordinated structure is the most preferable one, the same calculations have been performed for clusters of the type Li(S$_1$)$_n$(S$_2$)$_m^+$, where S$_1$=EC, S$_2$=DMC and n+m=4. The free energy changes for the transitions:

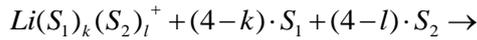

$$Li(S_1)_k(S_2)_l^+ + (4-k)\cdot S_1 + (4-l)\cdot S_2 \rightarrow$$

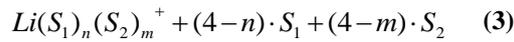

$$Li(S_1)_n(S_2)_m^+ + (4-n)\cdot S_1 + (4-m)\cdot S_2 \quad (3)$$

where $k+l=n+m=4$ and n,m,k,l=1-3, were also estimated, in order to find the most favorable structure among the clusters with two types of solvents around lithium.

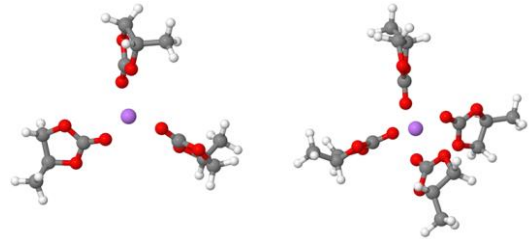

**Figure 2:** Structure of the optimized Li(PC)$_3^+$ and Li(PC)$_4^+$ clusters.

The calculated values of enthalpy, entropy and binding energy of the investigated Li(S$_1$)$_n$(S$_2$)$_m^+$ clusters, together with the free energy values of the Li(S$_1$)$_n$(S$_2$)$_m^+$ + (4-n) S$_1$ +(4-m) S$_2$ aggregates are presented in Tables 3,4. The results obtained reveal that among all the possible combinations of tetracoordinated clusters containing both EC and DMC the most preferable is Li(EC)$_3$(DMC)$^+$, followed by Li(EC)$_2$(DMC)$_2^+$ and eventually the Li(EC)(DMC)$_3^+$ is the least favorable among these three clusters. This finding is in agreement with both ab initio studies by Borodin and Smith[40], where the relative stabilities were based upon the energies of the complexes, and with the very recently reported study by Bhatt and O'Dwyer[25]. The structures of the optimized clusters are presented in Figure 4.

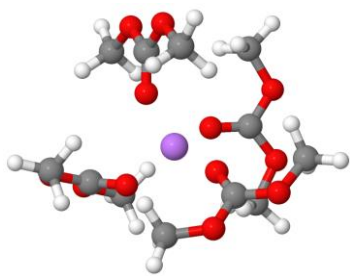

**Figure 3:** Structure of the optimized Li(DMC)$_4^+$ cluster.

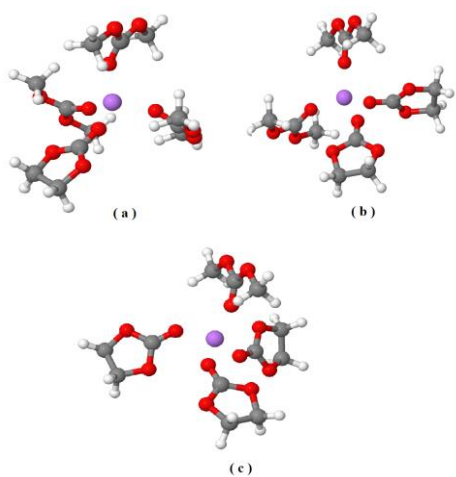

**Figure 4:** Structure of the optimized: (a) Li(EC)(DMC)$_3^+$, (b) Li(EC)$_2$(DMC)$_2^+$ and (c) Li(EC)$_3$(DMC)$^+$ clusters.

Finally, a similar analysis was performed for the clusters Li(S$_1$)$_n$(S$_2$)$_l$(S$_3$)$_m^+$ (S$_1$=EC, S$_2$=DMC, S$_3$=PC), containing EC, DMC and PC solvent molecules. The structures of the optimized clusters are presented in Figure 5.

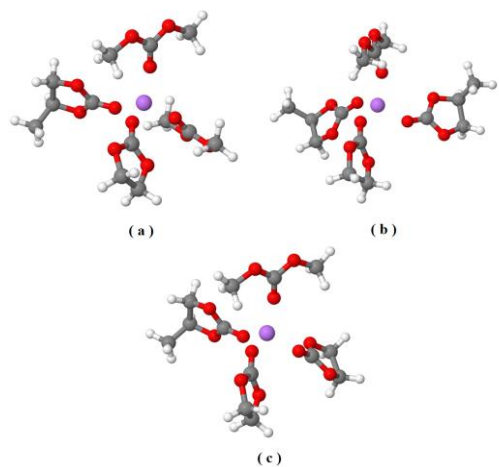

**Figure 5:** Structure of the optimized: (a) Li(EC)(PC)(DMC)$_2^+$, (b) Li(EC)(PC)$_2$(DMC)$^+$ and (c) Li(EC)$_2$(PC)(DMC)$^+$ clusters.

From the results obtained using the same methodology discussed above, which are presented in Tables 5,6, it can be seen that the cluster Li(EC)$_2$(DMC)(PC)$^+$, seems to be the most favorable one, followed by Li(EC)(DMC)(PC)$_2^+$ and finally by Li(EC)(DMC)$_2$(PC)$^+$. However the free energy changes for the transition paths between these clusters are very small, indicating that all these structures could possibly exist in the bulk phase. Also the free energy changes for the transition paths leading to the Li(EC)$_4^+$ and Li(PC)$_4^+$ clusters are very small. It has also been found that the transitions from clusters containing only EC and DMC to clusters with three types of solvent molecules, by substitution of one EC or DMC molecule with a PC one, are favorable in terms of the free energy changes. The substitution of a DMC molecule has been found to be more favorable than the substitution of an EC one. In general the above findings indicate that the presence of PC contributes to the stabilization of the clusters. This fact can be also supported by the calculated binding energies of the tetracoordinated clusters, which are plotted together in Figure 6.

**Table 3:** Thermodynamical parameters and Binding energies of $Li^+(EC)_n(DMC)_m$ (n+m=4) clusters (H and BE are in kcal/mol and S in cal mol$^{-1}$ K$^{-1}$)

| Clusters | H | BE | S |
|---|---|---|---|
| $Li^+(EC)_3(DMC)$ | -5632.2 | -116.3 | 182.1 |
| $Li^+(EC)_2(DMC)_2$ | -5790.4 | -113.0 | 186.2 |
| $Li^+(EC)(DMC)_3$ | -5946.2 | -109.1 | 204.9 |

H=Enthalpy, BE = Binding Energy, S = Entropy

**Table 4:** Gibbs free energies (in kcal/mol) of clusters containing two types of solvents.

| Clusters | G |
|---|---|
| $Li^+(EC)_3(DMC)$ + EC + 3 DMC | -11741.7 |
| $Li^+(EC)_2(DMC)_2$ + 2 EC + 2 DMC | -11737.3 |
| $Li^+(EC)(DMC)_3$ + 3 EC + DMC | -11735.0 |
| $Li^+(EC)_4$ + 4 DMC | -11744.2 |
| $Li^+(DMC)_4$ + 4 EC | -11729.4 |

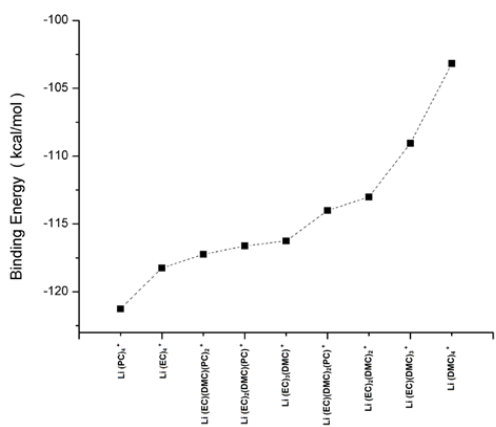

**Figure 6:** Calculated binding energies of the optimized tetracoordinated $Li^+$ clusters.

Interestingly, from Tables 4,6 it can also be seen that the presence of DMC contributes to the increase of the entropy and therefore to the decrease of the free energy of the clusters containing more than one solvent type. Therefore, although from an energetic point of view the preference of $Li^+$ to bind with the highly polar PC and EC molecules is undoubtful, from an entropic point of view the addition of DMC molecules contributes to the decrease of the free energy of the clusters. In the investigated finite size clusters the energetic contribution seems to be the most important one. However, the small free energy differences observed for the transition paths leading from one cluster to the other indicate that the formation of these clusters should be investigated at more extended length scales, in order to have a more realistic description of the effects taking place in the bulk liquid phase. This was the most important motivation to combine DFT calculations of isolated clusters with classical MD simulations at more extended length scales, in order to have a clearer picture about the solvation of $Li^+$ in pure and mixed organic carbonate-based solvents.

### 3.2 Liquid Solvation Structure

#### 3.2.1 Local Structural Inhomogeneities

By analyzing the trajectories obtained by the classical MD simulations described in section 2.2, the local structure around $Li^+$ were investigated for all the selected pure and mixed solvents. The local structure was analyzed in

**Table 5:** Thermodynamical parameters and Binding energies of $Li^+(EC)_l(DMC)_m(PC)_n$ (l+m+n=4) clusters (H and BE are in kcal/mol and S in cal mol$^{-1}$ K$^{-1}$)

| Clusters | H | BE | S |
|---|---|---|---|
| $Li^+(EC)_3(DMC)(PC)$ | -5999.3 | -116.6 | 189.0 |
| $Li^+(EC)(DMC)_2(PC)$ | -6157.0 | -114.0 | 202.5 |
| $Li^+(EC)(DMC)(PC)_2$ | -6367.2 | -117.2 | 188.9 |

H=Enthalpy, BE = Binding Energy, S = Entropy

**Table 6:** Gibbs free energies (in kcal/mol) of clusters containing three types of solvents.

| Clusters | G |
|---|---|
| $Li^+(EC)_2(DMC)(PC)$ + 2 EC + 3 PC + 3 DMC | -18781.2 |
| $Li^+(EC)(DMC)_2(PC)$ + 3 EC + 3 PC + 2 DMC | -18779.2 |
| $Li^+(EC)(DMC)(PC)_2$ + 3 EC + 2 PC + 3 DMC | -18780.3 |
| $Li^+(EC)_4$ + 4 DMC + 4 PC | -18783.3 |
| $Li^+(DMC)_4$ + 4 EC + 4 PC | -18768.5 |
| $Li^+(PC)_4$ + 4 EC + 4 DMC | -18781.7 |

terms of the most representative atom-atom pair radial distribution functions (prdf). The results obtained indicate a preferential interaction of Li$^+$ with the carbonyl oxygen atom (O$_C$) of the carbonate molecules, which is in agreement with all the previously reported experimental and theoretical studies. The calculated Li$^+$-O$_C$ prdfs between Li$^+$ and the carbonyl oxygen atoms of EC, PC and DMC in the pure, binary and ternary solvents are presented in Figure 7.

One of the main features of the calculated prdfs is the existence of a high intensity peak located at 1.8 Å in all cases. In the cases of the pure solvents the amplitude of this first peak decreases in the order $g(r)_{max}^{DMC} > g(r)_{max}^{PC} > g(r)_{max}^{EC}$. In the cases of the binary and ternary mixtures, the amplitude increases significantly in the case of DMC and decreases in the cases of EC and PC. The high amplitude of the first peak of the Li$^+$-O$_c$ prdf in the case of DMC indicates a strong interaction between Li$^+$ and the DMC molecules. The position of the first minimum, which determines the size of the first solvation shell, is located at about 2.6 Å in the case of the pure solvents and the binary mixture and at about 2.4 Å in the case of the ternary one. In order to estimate the number of solvent molecules around Li$^+$ in all the investigated systems, the local coordination numbers of each type of solvent were calculated as a function of the distance from the lithium cation and they are presented in Figure 8. The values of the peak positions and amplitudes of the calculated prdfs, together with the first minimum position and the corresponding coordination numbers for the first solvation shell are presented in Table 7.

From all these data it can be concluded that the solvation shell of Li$^+$ consists of four solvent molecules and this situation does not change in the cases of the binary and ternary solvent mixtures. At this point it should be mentioned that several MD simulation studies[27,28,40-47] of the solvation of Li$^+$ in pure carbonate solvents have given different estimations of the coordination number corresponding to the first solvation shell of Li$^+$. In particular, whereas some potential models predict a solvation shell of Li$^+$ which consists of four solvent molecules[27,28,40-42], other force fields predict a higher coordination number [43,45]. The potential models predicting higher coordination numbers also exhibit a strong dependence of the obtained coordination number on the salt concentration. From the shape of the obtained prdfs it is also clear that there are significant local structural inhomogeneities in the system. These structural fluctuations are reflected on the behavior of the prdfs, which reveal the presence of well-structured second and third coordination shells.

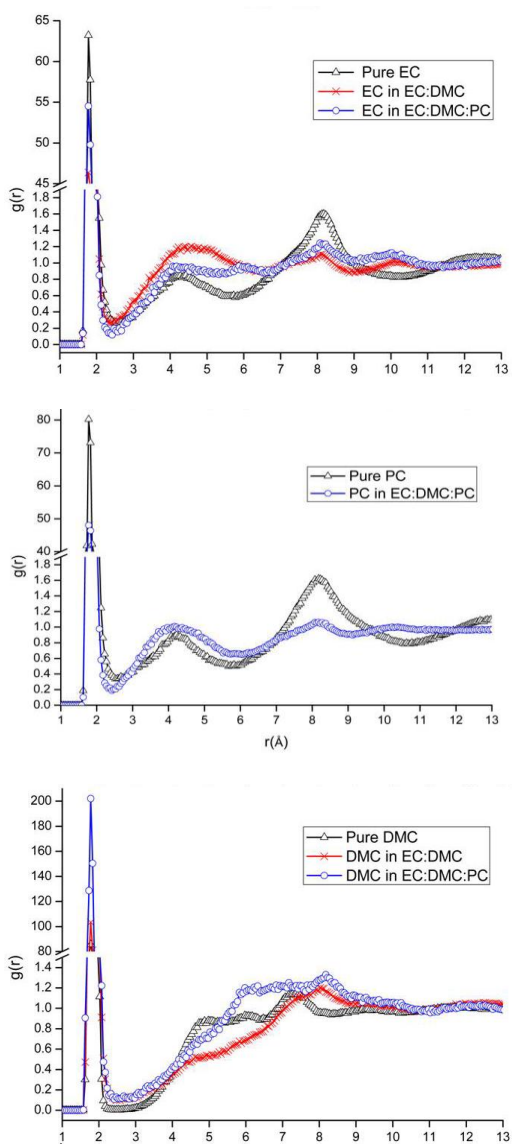

**Figure 7:** Calculated Li$^+$ - O$_c$ radial distribution functions.

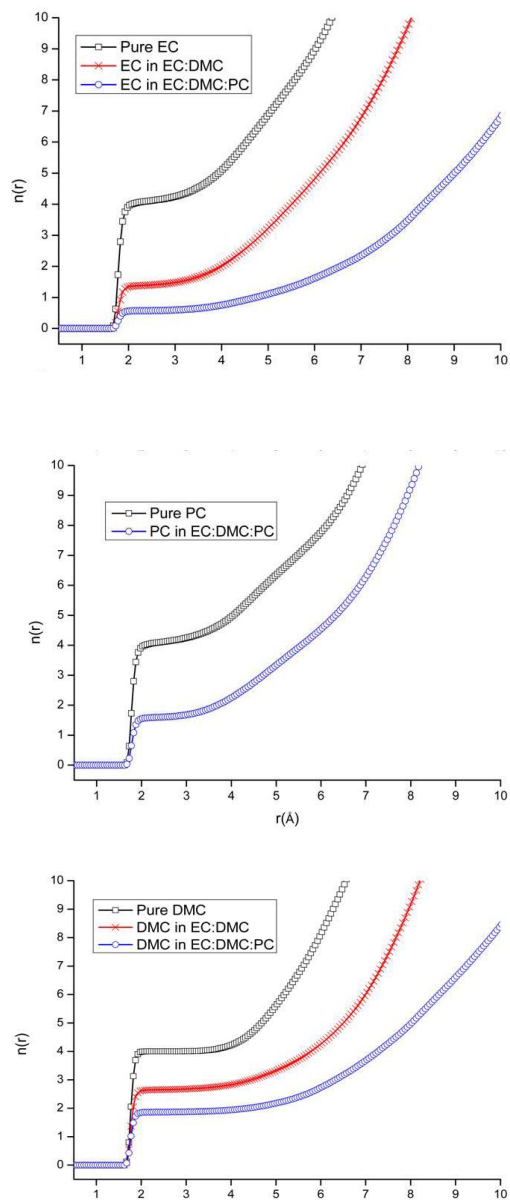

**Figure 8:** Calculated Li$^+$ - O$_c$ local coordination numbers.

Eventually they reach the homogeneous limit, obtaining values close to unity, only at length scales longer than 15 Å.

In the cases of mixtures of organic carbonate solvents only a limited number of studies[40,45,48] have focused on the solvation of Li$^+$ in binary mixtures. Also, to the best of our knowledge, this is the first simulation study devoted to the solvation of Li$^+$ in ternary mixtures of carbonate-based solvents. In binary and ternary mixtures, beyond the estimation of the coordination number of the primary solvation shell around lithium, the mixed composition of the shell is also of interest. The local compositional fluctuations in a spherical shell around Li$^+$ can be characterized in terms of the corresponding local mole fractions of each type of solvent[49]. For instance, in a ternary mixture of solvents with indices i, j, k the local mole fraction of solvent k around Li$^+$ is expressed as:

$$X_k(r) = \frac{n_k(r)}{n_i(r) + n_j(r) + n_k(r)} \quad (4)$$

**Table 7:** Peak positions and amplitudes of the calculated prdfs, together with the first minimum position and the corresponding coordination numbers for the first solvation shell of Li$^+$.

| System | EC | PC | DMC | EC:DMC (1:1) | EC:DMC:PC (1:1:3) |
|---|---|---|---|---|---|
| First Peak Position- Amplitude | 1.78 Å 63.19 | 1.78 Å 80.20 | 1.78 Å 85.58 | Li$^+$-EC: 1.78 Å 46.36 Li$^+$-DMC: 1.78 Å 103.29 | Li$^+$-EC: 1.78 Å 54.51 Li$^+$-DMC: 1.78 Å 202.09 Li$^+$-PC: 1.78 Å 48.03 |
| First Minimum Position | 2.58 Å | 2.53 Å | 2.58 Å | Li$^+$-EC: 2.53 Å Li$^+$-DMC: 2.58 Å | Li$^+$-EC: 2.43 Å Li$^+$-DMC: 2.43 Å Li$^+$-PC: 2.43 Å |
| Coordination Number (First Shell) | 4.13 | 4.12 | 4.00 | Li$^+$-EC: 1.41 Li$^+$-DMC: 2.66 | Li$^+$-EC: 0.58 Li$^+$-DMC: 1.86 Li$^+$-PC: 1.60 |

In Eq. 4, $n_k(r)$ is the coordination number of solvent k corresponding to a spherical shell of radius r around Li$^+$. In a similar way, all the local mole fractions for the different types of solvents in the binary and ternary mixture around Li$^+$ have been calculated and are presented in Figure 9. The coordination number considered in the present calculations corresponds to the carbonyl oxygen atoms of each solvent around Li$^+$. The results obtained from the present MD simulations actually reveal that in the cases of binary and ternary mixtures there is a high concentration of DMC in the first solvation shell of Li$^+$. The existence of significant compositional fluctuations around Li$^+$ can also be observed. As a result of these fluctuations, the structure around Li$^+$ could be described in terms of a short-range structure (up to 4-5 Å from Li$^+$) where the composition is rich in DMC, and by a longer-range structure where in the case of the binary solvent mixture the composition is rich in EC and, in the case of the ternary one, it is rich in PC. At larger length scales, in the range of about 15 Å from the lithium cation, the local mole fractions of each solvent reach the corresponding bulk value. From Figure 9 it can also be observed that the typical length scale of these fluctuations around Li$^+$ is slightly more extended in the case of the ternary mixture than in the binary one.

The findings of the MD simulations regarding the composition of the first solvation shell of Li$^+$ in binary and ternary mixtures could be in contradiction with the results obtained by the DFT calculations.

the first solvation shell, the use of larger size clusters in DFT calculations seems to be needed.

To obtain a more detailed picture of the multiple different local microstructures observed in the first solvation shell of Li$^+$, an analysis of the different clusters Li(S$_1$)$_n$(S$_2$)$_l$(S$_3$)$_m^+$ (S$_1$=EC, S$_2$=DMC, S$_3$=PC, n,l,m= 0 – 4 and n+l+m=4) observed inside this shell was performed for the binary and ternary solvent mixtures. The relative populations of these are shown in Figure 10.

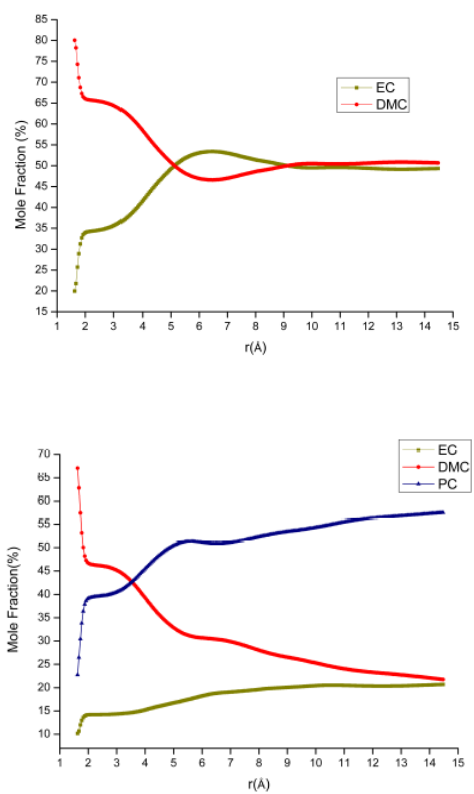

**Figure 9:** Local mole fractions (%) of EC, PC, DMC as a function of the distance from the lithium cation in the binary and ternary mixtures.

However, it should be pointed out that the first shell structure predicted in the MD simulations could be the result of local structural correlations between the first, second and third solvation shells which exhibit completely different compositions. Indeed, the collective effects arising due to the interactions between regions of different density and composition could possibly stabilize the local structure observed inside the first solvation shell of Li$^+$. These are clearly not taken into account in the cases of the DFT calculations, where isolated clusters have been investigated. To perform a meaningful comparison between MD and DFT results in cases where significant structural fluctuations are present, extended far beyond

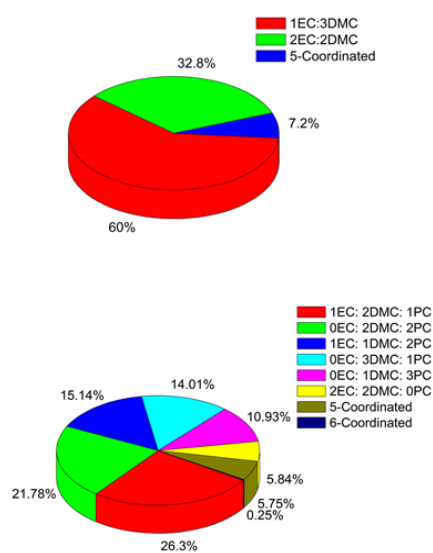

**Figure 10:** Fractions of the Li(EC)$_n$(DMC)$_{4-n}^+$ and Li(EC)$_n$(PC)$_m$(DMC)$_{4-n-m}^+$ clusters in the binary and ternary mixture.

In the binary mixture, the most prominent clusters inside the first solvation of Li$^+$ are the Li(EC)(DMC)$_3^+$ and the Li(EC)$_2$(DMC)$_2^+$, with the first exhibiting almost double occurrence probability. There is also a small fraction of clusters (7.2 %) where Li$^+$ is pentacoordinated. About 80 % of these 5-coordinated Li$^+$ complexes consist of 2 EC and 3 DMC molecules, 19 % of 3 EC and 2 DMC molecules and an almost negligible 1 % fraction consists of 1 EC and 4 DMC molecules. In the ternary mixture the most prominent clusters are Li(EC)(DMC)$_2$(PC)$^+$ and Li (DMC)$_2$(PC)$_2^+$ , having though non-negligible fractions of Li (EC)(DMC)(PC)$_2^+$, Li (DMC)$_3$(PC)$^+$, Li (DMC)(PC)$_3^+$ and a smaller one corresponding to the Li(EC)$_2$(DMC)$_2^+$ cluster. It should be noted that almost half of the observed clusters with four solvent molecules (49.7 %) do not contain EC at all, whereas DMC is involved in the formation of all of them. PC is absent in only 6.2 % of the total fraction of clusters with four solvent molecules. There is also a fraction 5.75 % of clusters with five solvent molecules and an almost negligible one (0.25 %) of clusters where Li$^+$ is hexacoordinated.

Interestingly, such a discrepancy between the results obtained by MD and DFT calculations has been also reported by Borodin and Smith[40]. In their study, Borodin and Smith performed a MD simulation of a solution of LiPF$_6$ in an equimolar EC-DMC binary mixture (EC+DMC:Li=11.8) using a polarizable potential model. The coordination number of DMC in the first solvation shell of Li$^+$ was higher than the coordination number of EC. This is in contrast with the results reported by Tenney and Cygan[43] , favoring the presence of EC in the first solvation shell of Li$^+$ in a low concentration LiPF$_6$ solution in a binary equimolar EC-DMC mixture. However, there is a significant difference between the potential models employed in these two MD simulation studies. The potential model used by Borodin and Smith[40] mainly predicts tetracoordinated Li$^+$ complexes inside its first solvation shell, whereas in the case of the simulation reported by Tenney and Cygan[45] the solvation shell of Li$^+$ mainly consists of five or six solvent molecules. The potential model used in the presented study also favors local structures with four solvent molecules around Li$^+$. This observation was an additional motivation to search for possible correlations between the formation of a local tetrahedral structure and the preference of Li$^+$ to form a first solvation shell mainly consisting of DMC molecules.

### 3.2.2 First solvation shell orientational ordering

The local tetrahedral structure around lithium can be investigated in terms of the tetrahedral order parameter, q$^{50}$. This parameter q provides information about the extent to which a particle and its four nearest neighbors adopt a tetrahedral arrangement and is defined as:

$$q = 1 - \left\langle \frac{3}{8}\sum_{j=1}^{3}\sum_{k=j+1}^{4}\left(\cos\phi_{jik} + \frac{1}{3}\right)^2 \right\rangle \quad (5)$$

In this equation $\phi_{jik}$ corresponds to the angle formed by the vectors $\vec{r}_{ij}$ and $\vec{r}_{ik}$, connecting the lithium cation i with the oxygen atoms of two of its four nearest neighbours j, k. Using this definition, q=1 in a perfect tetrahedral network and q=0 in an ideal gas[48]. Kumar et al[51] have also suggested that an entropic term associated with this local tetrahedral order around a particle can be calculated from the probability distribution of the tetrahedral order parameter P(q):

$$S_{tetr} = \frac{3}{2} \cdot K_B \int \ln\left(1-q\right) \cdot P(q) \cdot dq \quad (6)$$

We note that in general, the entropy of a fluid can be expressed in terms of several contributions arising from the integrals over multiparticle correlation functions[52-54]. However, our analysis revealed that orientational entropy plays a crucial role in the stabilization of the local solvation shell structures, and hence we have focused on this particular aspect of orientational order and entropy. Also Kumar et al [51] have proposed that in systems exhibiting significant local tetrahedral orientational order, the tetrahedral entropy captures the most important contributions arising from configurational orientational correlations[54] and therefore offers a rather simpler way to investigate the entropy associated with such kind of local structures.

The normalized distributions P(q) have been calculated for all the simulated systems and are presented in Figure 11.

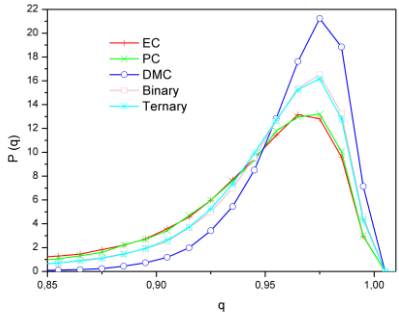

**Figure 11:** Normalized distribution of the tetrahedral order parameter around $Li^+$ in all the investigated solvents.

The average values of q and $S_{tetr}$ are shown in Table 8. From the shape of these distributions it can be seen that there is a very significant tetrahedral local ordering around $Li^+$ in all cases and the addition of cosolvents does not significantly distort this structural order. The average values of q are in the range of about 0.93-0.96 in all cases, very close to the ideal tetrahedron value. Whereas there is no significant difference in the average values of q, the calculated value of $S_{tetr}$ is noticeably lower in the case of the pure DMC solution, signifying that from an entropic point of view the presence of DMC contributes to the stabilization of the local tetrahedral structure around $Li^+$.

As it has been pointed out by one of the Authors in a previous publication[55], given the additive property of the integral in Eq. 6, the contributions of different cluster structures to the calculated $S_{tetr}$ value can be easily extracted. By using the fractions $\chi_{nlm}$ of the different cluster structures $Li(S_1)_n(S_2)_l(S_3)_m^+$ ($S_1$=EC, $S_2$=DMC, $S_3$=PC), already shown in Figure 10, the tetrahedral entropic term $S_{tetr}$ per lithium cation is given by $S_{tetr} = \sum_{nlm} \chi_{nlm} \cdot S_{tetr}^{nlm}$, where the contribution of each cluster state is:

$$S_{tetr}^{nlm} = \frac{3}{2} \cdot K_B \int \ln(1-q) \cdot P^{nlm}(q) \cdot dq \quad (7)$$

In this equation $P^{nlm}(q)$ is the normalized probability density distribution of the tetrahedral order parameter q corresponding to a cluster $Li(S_1)_n(S_2)_l(S_3)_m^+$ inside the first solvation shell of $Li^+$. The calculated values of $S_{tetr}^{nlm}$ corresponding to all the tetracoordinated $Li^+$ clusters observed during the simulations of the dilute solutions of $Li^+$ in the pure, binary and ternary solvents are presented in Figure 12. In general it can be observed that the clusters having a high fraction of DMC molecules (3 and 4 molecules) exhibit a higher tetrahedrality, which is reflected in the lower values of $S_{tetr}^{nlm}$. Interestingly there is also a small difference in the $S_{tetr}^{nlm}$ values of the clusters Li(EC)$_2$(DMC)$_2^+$ observed in the binary and ternary solvent mixtures, the Li(EC)$_2$(DMC)$_{2\,(b)}^+$ and Li(EC)$_2$(DMC)$_{2\,(t)}^+$ clusters, respectively. This is a clear indication of the effect of the second solvation shell structure, which is different in the binary and ternary solvent mixtures, on the stabilization of the local structure observed in the first solvation shell. This observation further supports our previous statements that the use of larger size clusters in DFT calculations is required in cases where there are significant structural fluctuations, extended far beyond the first solvation shell.

**Table 8:** Average values of q and $S_{tetr}/K_B$ per lithium cation obtained by the MD simulations.

| System | EC | PC | DMC | EC:DMC (1:1) | EC:DMC:PC (1:1:3) |
|---|---|---|---|---|---|
| $<q>$ | 0.934 | 0.932 | 0.964 | 0.948 | 0.947 |
| $S_{tetr}/K_B$ | -4.56 | -4.52 | -5.29 | -4.87 | -4.84 |

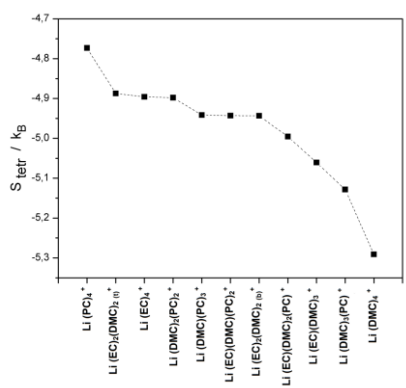

**Figure 12:** The calculated values of $S_{tetr}^{nlm}$ corresponding to all the tetracoordinated Li$^+$ clusters observed in the pure, binary and ternary solvents.

It should be mentioned that previous experimental studies, using electrospray ionization and mass spectrometry (ESI-MS) techniques[56], have predicted a stronger preference of Li$^+$ to bind with EC than with DMC in binary EC-DMC solvents. This preference has been attributed to the stronger ion dipole interactions between Li$^+$ and EC molecules. However, the Authors in that study pointed out that their results have been extracted from measurements of isolated clusters formed by two solvent molecules only instead of four. This difference has been attributed to the relative depletion of free solvent molecules at high salt concentrations or to a possible partial desolvation, during which the loosely bound molecules in the Li(S)$_4^+$ clusters (S: solvent) could be lost before the solvated species ever reach the aperture of the spectrometer.

Borodin and Smith[40,57] pointed out in their study, where they observed a slight preference for Li$^+$ to bind with DMC, that the composition of the Li$^+$ solvation can be different in the gas and liquid phases. Takeuchi et al.[42] have come to similar conclusions about the Li$^+$ binding patterns to a PF$_6^-$ anion. Another issue, which has not been discussed in the previous ESI-MS investigation, is the importance of the presence of the anions in the stabilization of the Li$^+$ solvation shell. The investigated solutions had a high salt concentration and therefore the presence of the anions could strongly affect the preferential binding of Li$^+$ with the solvent molecules. The preferential solvation around Li$^+$ in binary and ternary mixtures is a quite complicated issue and besides the relative permittivity and donicity of the cyclic and acyclic components of the mixed solvent[56], salt concentration possibly plays a very important role. This statement is also supported by the findings of Borodin and Smith[40], where the presence of $PF_6^-$ anions in the solvation shell around $Li^+$ strongly affects the preferential binding of $Li^+$ with the cyclic and acyclic solvent molecules. This was the main reason why in the presented study dilute solution of $Li^+$ were selected to be investigated, to focus more on the effects arising just due to the individual cation-solvent and solvent-solvent interactions. However, the anion concentration effect on the local solvation structure of mixed electrolyte solvents is a very important topic and will be the subject of a forthcoming study.

In general the preferential solvation around $Li^+$ is being discussed in terms of the strength of individual pair cation-dipole interactions. In this sense, the observed preferential solvation of $Li^+$ by the non-polar DMC molecules instead of the highly polar EC and PC molecules seems to be problematic. To obtain a deeper insight, having in mind that collective effects might play a very important role in the solvation phenomena in condensed phases, the total dipole of the solvent molecules inside the solvation shell of $Li^+$ was calculated as:

$$\vec{M}_{sh} = \sum_i \vec{\mu}_i \text{, with } r_{i-Li} \leq r_c \quad (8)$$

These local collective dipole moments, corresponding to the sum of the individual molecular dipoles of the solvent molecules which are inside the first solvation shell of $Li^+$, have been extracted by analyzing the trajectories of the classical MD simulations.

It should be also noted that although the effects arising from induction interactions are not taken into account in the classical simulations, in contrast with *ab initio* methods, the atomic charges and molecular geometries used in the classical force fields have been extracted from high level *ab initio* calculations[27,58] and predict dipole moments which are slightly higher than those reported experimentally for the gas phase. This is a very common approach to build effective classical potential models for liquid simulations, since due to polarization effects the dipole moments in the liquid phase are slightly higher than the gas phase ones. The effective potential models used also predict with a very good accuracy a wide range of properties of the neat liquid solvents[27,58].

In Eq. 8, $r_c = 2.6$ Å is the distance determining the size of the solvation shell taken into account in the calculations. The calculated normalized probability density distributions $P(M_{sh})$ of the magnitude of the total dipole $M_{sh} = |\vec{M}_{sh}|$ for all the investigated systems are presented in Figure 13.

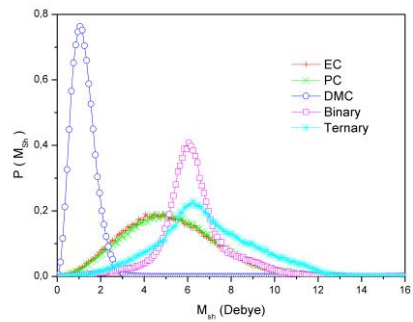

**Figure 13:** The calculated normalized probability density distributions $P(M_{sh})$ of the magnitude of the total dipole $M_{sh} = |\vec{M}_{sh}|$ inside the solvation shell of $Li^+$.

From these data it is clear that inside the first solvation shell of $Li^+$ in ternary and binary mixtures, which exhibits a high concentration of non-polar DMC molecules, the total dipole of the solvent molecules is higher than in the cases where $Li^+$ is solvated by the highly polar EC and PC molecules. This finding signifies that the observed tetrahedral arrangement of the solvent molecules inside the solvation shell of Li$^+$ causes a cancellation of the individual molecular dipole vectors and this cancellation is more important in the cases where molecules of the same geometry are present. When different types of solvent molecules are tetrahedrally packed inside the first solvation the resulting total dipole is higher, even though the fraction of the non-polar DMC molecules is higher. As a result, the local environment around Li$^+$ becomes more polar when the fraction of DMC is higher. This observation clearly indicates that collective effects are very important in determining the local dielectric environment around a Li$^+$ cation and the relative binding preferences should be determined on the basis of this collective dipole moment of the solvation shell and not in terms of pair ion-dipole interactions.

Before closing this section it would be useful to point out again that the experimental determination of the coordination number around the lithium ions in pure carbonate-based solvents has not been definitely resolved[7-18]. As it was mentioned in the introduction, it should be however emphasized that designing experimental methods or theoretical models to analyze the experimental data in order to provide a direct measurement of the coordination number is an extremely complicated task[7,19,20]. Nevertheless, recent advances[59] in coupling theoretical methods with experiments to bias molecular simulations with experimental data tend to favor a tetrahedral-like structure, as also sophisticated polarizable models[40], which predict a wide range of properties very close to the experimental ones. This was the main reason why a simple model, predicting a tetrahedral network around Li$^+$, was selected in the present treatment. In general, one of the most reliable experimental methods to extract information about the local structure in liquids is neutron diffraction. Only one neutron diffraction study of a 10 mol % solution of LiPF$_6$ in liquid PC has been reported up to now[12]. In this study the authors, by integrating the total radial distribution function G(r) obtained by the diffraction experiments up to its first minimum, have estimated the coordination number around Li$^+$ to be 4.5. However, they have assumed that all the contributions to the total G(r) at the short-range part up to 2.4 Å arise from the Li$^+$-O$_c$ prdf. By inspecting the above mentioned theoretical studies[40,59] and by performing a trial simulation of the same system, it became obvious that the Li$^+$-F prdf contributes to the shape of the total G(r) at this short-range. Therefore, a part of this 4.5 coordination number arises from the Li$^+$-F coordination number. But the cation-anion coordination number at the same distance range is definitely smaller, since the Li$^+$-F prdf and the corresponding coordination number provide information about the number of fluorine atoms inside the solvation shell of Li$^+$, which of course could belong to the same PF$_6^-$ anion. As a consequence, it is very likely that the coordination number giving the total number of solvent molecules and anions around Li$^+$ should be smaller and could be even closer to four, or less. However, from that neutron diffraction study it's not possible to extract the separate contributions of the Li$^+$-O$_c$ and Li$^+$-F prdfs in order to have a more accurate estimation of the total coordination number in the first solvation shell of Li$^+$. From this observation it becomes clear that even the interpretation of experimental data should be treated very carefully and if it is also combined with reverse modeling techniques[60] it could provide even more valuable information.

**4. Concluding Remarks**

In the present study the solvation of $Li^+$ in pure carbonate-based solvents and their binary and ternary mixtures has been investigated, using a combination of quantum chemical DFT calculations and classical MD simulations. An analysis based on the changes in the free energy associated to the transitions between different types of clusters formed by the lithium cation and the solvent molecules has shown that the formation of complexes with four solvent molecules is the most favorable. In the cases of clusters containing EC and DMC molecules, the DFT calculations have predicted a preferential binding with EC rather than with DMC. For the clusters including at the same type EC, PC and DMC molecules, although the $Li(EC)_2(DMC)(PC)^+$ has been predicted to be the most favorable one, the free energy changes for the transition paths between several clusters are very small, indicating that all these structures could possibly exist in the bulk phase.

On the other hand the MD simulations have revealed the existence of significant structural heterogeneities, extending up to a length scale which is more than five times the size of the first shell radius. These heterogeneities are very important in the cases of the binary and ternary solvent mixtures, revealing very significant compositional inhomogeneities around a lithium cation. As a result of these fluctuations, the structure around Li+ in the mixed solvents can be described in terms of a short-range structure (up to 4-5 Å from Li+), where the composition is rich in DMC, and by a longer-range structure where in the case of the binary solvent mixture the composition is rich in EC and in the case of the ternary one it is rich in PC. The local mole fractions of each solvent obtain their bulk values at more extended length scales, as it was mentioned above. This finding indicates that the collective effects arising due to the interactions between regions of different composition could possibly stabilize the local structure of the first solvation shell of $Li^+$.

In the DFT calculations of the finite size isolated clusters, although from an energetic point of view the preference of $Li^+$ to bind with the highly polar PC and EC molecules is undoubtful, from an entropic point of view the addition of DMC molecules contributes to the decrease of the free energy of the clusters. In these finite size clusters the energetic contribution seems to be the most important one. However, the small free energy differences observed for the transition paths leading from one cluster to the other indicate that the formation of these clusters should be investigated at more extended length scales in order to have a more realistic description of the effects taking place in the bulk liquid phase. The existence of significant structural fluctuations in the bulk liquid phases, extended far beyond the first solvation shell, signifies that to achieve a direct comparison between the MD and the DFT results the use of larger size clusters in DFT calculations seems to be more appropriate.

The MD simulations have also revealed that there is a very significant tetrahedral local ordering around $Li^+$ in all cases and the addition of cosolvents does not distort this structural order. By calculating an entropic term associated with this local tetrahedral order around Li$^+$, it has also been revealed that the clusters having a high fraction of DMC molecules (3 and 4 molecules) exhibit a higher tetrahedrality, which is reflected in the lower values of this entropic term. Therefore, from an entropic point of view the presence of DMC contributes to the stabilization of the local tetrahedral structure around Li$^+$.

A very interesting finding also revealed in the present study is that inside the first solvation shell of Li$^+$ in the ternary and binary mixtures, which exhibit a high concentration of non-polar DMC molecules, the total dipole of the solvent molecules is higher than in the cases where Li$^+$ is solvated by the highly polar EC and PC molecules. The observed local tetrahedral packing of the solvent molecules in the first solvation shell of Li$^+$ causes a cancellation of the individual molecular dipole vectors, which seems to be more important in the cases where molecules of the same type are present. These collective effects are very important in determining the local permittivity around a Li$^+$ cation and the relative binding preferences should be based upon this total dipole moment of the solvation shell and not in terms of the pair ion-dipole interactions, which is traditionally considered as the most important factor determining the preferential solvation in such systems. Further work is in progress to elucidate the impact of the above static structural features on dynamic and transport properties of the investigated electrolytes.

**Acknowledgements:**

The authors acknowledge financial support from ANR-2011 PRGE002-04 ALIBABA and DSM Energie CEA Program.

**Supporting Information Available:**

The details of the inter- and intra-molecular classical force fields employed in the present study are provided in the Supporting Information as a DL-POLY FIELD file.

**References**

1) Xu, K. Nonaqueous Liquid Electrolytes for Lithium-based Rechargeable Batteries. *Chem. Rev.* **2004**, *104*, 4303

2) Armand, M.; Tarascon, J. M. Building Better Batteries. *Nature* **2008**, *541*, 652

3) Kim, Y.; Goodenough, J. B. Challenges for Rechargeable Li Batteries. *Chem. Mater.* **2010**, *22*, 587

4) Besenhard, J.O.; Winter, M.; Yang, J.; Biberacher, W. Filming Mechanism of Lithium-Carbon Anodes in Organic and Inorganic Electrolytes. *J. Power Sources* **2005**, *54*, 228

5) Yuan, K.; Bian, H.; Shen, Y.; Jiang, B.; Li, J.; Zhang, Y.; Chen, H.; Zheng, J. Coordination Number of Li$^+$ in Nonaqueous Electrolyte Solutions Determined by Molecular Rotational Measurements. *J. Phys. Chem. B* **2014**, *118*, 3689

6) Eshetu, G. G.; Bertrand, J.P.; Lecocq, A.; Grugeon, S.; Laruelle, S.; Armand, M.; Marlair, G. Fire behavior of carbonates-based electrolytes used in rechargeable batteries with a focus on the role of the LiPF$_6$ and LiFSI salts. *J. Power Sources* **2014**, *269*, 804

7) Bogle, X.; Vazquez, R.; Greenbaum, S.; von Wald Cresce, A.; Xu, K. Understanding Li$^+$- Solvent Interaction in Nonaqueous Carbonate Electrolytes with $^{17}$O NMR. *J. Phys. Chem. Lett.* **2013**, *4*, 1664-1668

8) Castriota, M.; Cazzanelli, E.; Nicotera, I.; Coppola, L.; Oliviero, C.; Ranieri, G.A. Temperature dependence of lithium ion solvation in ethylene carbonate-LiClO$_4$ solutions. *J. Chem. Phys.* **2003**, *118*, 5537-5541

9) Yuan, K.; Bian, H.; Shen, Y.; Jiang, B.; Li, J.; Zhang, Y.; Chen, H.; Zheng, J. Coordination Number of Li$^+$ in Nonaqueous Electrolyte Solutions Determined by Molecular Rotational Measurements. *J. Phys. Chem. B* **2014**, *118*, 3689-3695

**10)** Xu, K.; Lam, Y.; Zhang, S.S.; Jow, T. R.; Curtis, T.B. Solvation Sheath of Li$^+$ in Nonaqueous Electrolytes and Its Implication of Graphite/Electrolyte Interface Chemistry. *J. Phys. Chem. C* **2007**, *111*, 7411-7421

**11)** Smith, J. W.; Lam, R.K.; Sheardly, A.T.; Shih, O.; Rizzuto, A.M.; Borodin, O.; Harris, S.J.; Prendergast, D.; Saykally, R.J. X-Ray absorption spectroscopy of LiBF$_4$ in propylene carbonate: a model lithium ion battery electrolyte. *Phys. Chem. Chem. Phys.* **2014**, DOI: 10.1039/c4cp03240c

**12)** Kameda, Y.; Umebayashi, Y.; Takeuchi, M.; Wahab, M.A.; Fukuda, S.; Ishiguro, S.; Sasaki, M.; Amo, Y.; Usuki, T. Solvation structure of Li$^+$ in concentrated LiPF$_6$-propylene carbonate solutions. *J. Phys. Chem. B* **2007**, *111*, 6104-6109

**13)** Morita, M.; Asai, Y.; Yoshimoto, N.; Ishikawa, M. A Raman spectroscopic study of organic electrolyte solutions based on binary solvent systems of ethylene carbonate with low viscosity solvents which dissolve different lithium salts. *J. Chem. Soc. Faraday Trans.* **1998**, *94*, 3451-3456

**14)** Tsunekawa, H.; Narumi, A.; Sano, M.; Hiwara, A.; Fujita, M.; Yokoyama, H. Solvation and Ion Association Studies of LiBF$_4$-Propylenecarbonate-Trimethyl Phosphate Solutions. *J. Phys. Chem. B* **2003**, *107*, 10962-10966

**15)** Nie, M.; Abraham, D.P.; Seo, D.M.; Chen, Y.; Bose, A.; Lucht, B.L. Role of Solution Structure in Solid Electrolyte Interphase Formation on Graphite with LiPF$_6$ in Propylene Carbonate. *J. Phys. Chem. C* **2013**, *117*, 25381-25389

**16)** Li, Y.; Xiao, A.; Lucht, B.L. Investigation of solvation in lithium ion battery electrolytes by NMR spectroscopy. *J. Mol. Liq.* **2010**, *154*, 131-133

**17)** Kondo, K.; Sano, M.; Hiwara, A.; Takehiko, O.; Fujita, M.; Kuwae, A.; Iida, M.; Mogi, K.; Yokoyama, H. Conductivity and Solvation of Li$^+$ Ions of LiPF$_6$ in Propylene Carbonate Solutions. *J. Phys. Chem. B* **2000**, *104*, 5040-5044

**18)** Barthel, J.; Buchner, R.; Wismeth, E. FTIR Spectroscopy of Ion Solvation of LiClO$_4$ and LiSCN in Acetonitrile, Benzonitrile and Propylene Carbonate. *J. Solution Chem.* **2000**, *29*, 937-954

**19)** Buchner, R. Dielectric Spectroscopy of Solutions. In *Novel Approaches to the Structure and Dynamics of Liquids: Experiments, Theories and Simulations*; Samios J.; Durov, V.A. Eds.; Kluwer Academic Publishers; 2004; p 265.

**20)** Ravel, B.; Kelly, S.D. The Difficult Chore of Measuring Coordination by EXAFS *AIP Conf. Proc.* **2007**, *882*, 150-152

**21)** te Velde, G.; Bickelhaupt, F.M.; van Gisbergen, S.J.A.; Fonseca Guerra, C.; Baerends, E.J.; Snijders, J.G.; Ziegler, T. Chemistry with ADF. *J. Comput. Chem.* **2001**, *22*, 931-967

**22)** Perdew, J.P.; Burke, K.; Ernzerhof, M. Generalized gradient approximation made simple. *Phys. Rev. Lett.* **1996**, *77*, 3865-3868

**23)** Boys, S.F.; Bernardi, F. Calculation of Small Molecular Interactions by Differences of Separate Total Energies – Some Procedures with Reduced Errors. *Mol. Phys.* **1970**, *19*, 553

**24)** Bhatt, M.D.; Cho, M.; Cho, K. Interaction of Li$^+$ ions with ethylene carbonate (EC): Density Functional Theory calculations. *Appl. Surf. Sci.* **2010**, *257*, 1463-1468

**25)** Bhatt, M.D.; O'Dwyer, C. Density functional theory calculations for ethylene carbonate-based binary electrolyte mixtures in lithium ion batteries. *Curr. Appl. Phys.* **2014**, *14*, 349-354

**26)** Martínez, L.; Andrade, R.; Birgin, E.G.; Martínez, J.M. PACKMOL: A Package for Building Initial Configurations for Molecular Dynamics Simulations. *J. Comput. Chem.* **2009**, *30*, 2157-2164

**27)** Soetens, J.C.; Millot, C.; Maigret, B. Molecular Dynamics Simulation of Li$^+$BF$_4^-$ in Ethylene Carbonate, Propylene Carbonate and Dimethyl Carbonate Solvents. *J. Phys. Chem. A* **1998**, *102*, 1055-1061

**28)** Masia, M.; Probst, M.; Rey, R. Ethylene Carbonate-Li$^+$: A Theoretical Study of Structural and Vibrational Properties in Gas and Liquid Phases. *J. Phys. Chem. B* **2004**, *108*, 2016-2027

**29)** Katon, J.E.; Cohen, M.D. Conformational Isomerism and Oriented Polycrystal Formation of Dimethyl Carbonate. *Can. J. Chem.* **1974**, *52*, 1994-1996

**30)** Katon, J.E.; Cohen, M.D. The Vibrational Spectra and Structure of Dimethyl Carbonate and its Conformational Behavior. *Can. J. Chem.* **1975**, *53*, 1378-1386

**31)** Thiebaut, J.M.; Rivail, J.L.; Greffe, J.L. Dielectric Studies of non Electrolyte Solutions. *J. Chem. Soc. Faraday Trans.2* **1976**, *72*, 2024-2034

**32)** Smith, W.; Forester, T. R. Parallel macromolecular simulations and the replicated data strategy: II. The RD-SHAKE algorithm. *Comput. Phys. Commun.* **1994**, *79*, 63-77

**33)** Ryckaert, J. P.; Ciccotti, G.; Berendsen, H. J. C. Numerical integration of the Cartesian equations of motion of a system with constraints: molecular dynamics of n-alkanes. *J. Comp. Phys.* **1977**, *23*, 327-341

**34)** Allen, M. P.; Tildesley, D. J., *Computer Simulations of Liquids,* Oxford University Press, Oxford, 1987

**35)** Hoover, W. G. Canonical dynamics: Equilibrium phase-space distributions. *Phys. Rev. A* **1985**, *31*, 1695-1697

**36)** Hoover, W. G. Constant pressure equations of motion. *Phys. Rev. A* **1986**, *34*, 2499-2500

**37)** Smith, W.; Forester, T. R. DL_POLY_2.0: A general-purpose parallel molecular dynamics simulation package. *J. Mol. Graphics* **1996**, *14*, 136-141

**38)** McQuarrie, D.A.; Simon, J.D. *Molecular Thermodynamics*; University Science Books; 1999

**39)** Swart, M.; Rösler, E.; Bickelhaupt, F.M. Proton Affinities of Maingroup-Element Hydrides and Noble Gases: Trends Across the Periodic Table, Structural Effects, and DFT Validation. *J. Comp. Chem.* **2006**, *27*, 1486

**40)** Borodin, O.; Smith, G.D. Quantum Chemistry and Molecular Dynamics Simulation Study of Dimethyl Carbonate: Ethylene Carbonate Electrolytes Doped with $LiPF_6$. *J. Phys. Chem. B* **2009**, *113*, 1763-1776

**41)** Takeuchi, M.; Kameda, Y.; Umebayashi, Y.; Ogawa, S.; Sonoda, T.; Ishiguro, S.; Fujita, M.; Sano, M. Ion-ion interactions of $LiPF_6$ and $LiBF_4$ in propylene carbonate solutions. *J. Mol. Liq.* **2009**, *148*, 99-108

**42)** Takeuchi, M.; Matubayashi, N.; Kameda, Y.; Minofar, B.; Ishiguro, S.; Umebayashi, Y. Free-Energy and Structural Analysis of Ion Solvation and Contact Ion-Pair Formation of $Li^+$ with $BF_4^-$ and $PF_6^-$ in Water and Carbonate Solvents. *J. Phys. Chem. B* **2012**, *116*, 6476-6487

**43)** Postupna, O.O.; Kolesnik, Y.V.; Kalugin, O.N.; Prezhdo, O.V. Microscopic Structure and Dynamics of $LiBF_4$ Solutions in Cyclic and Linear Carbonates. *J. Phys. Chem. B* **2011**, *115*, 14563-14571

**44)** Bhatt, M.D.; Cho, M.; Cho, K. Density Functional theory calculations and ab initio molecular dynamics simulations for diffusion of $Li^+$ within liquid ethylene carbonate. *Modelling Simul. Mater. Sci. Eng.* **2012**, *20*, 065004

**45)** Tenney, C.M.; Cygan, R.T. Analysis of Molecular Clusters in Simulations of Lithium-Ion battery Electrolytes. *J. Phys. Chem. C* **2013**, *117*, 24673-24684

**46)** Jorn, R.; Kumar, R.; Abraham, D.P.; Voth, G.A. Atomistic Modeling of the Electrode-Electrolyte Interface in Li-Ion Energy Storage Systems: Electrolyte Structuring. *J. Phys. Chem. C* **2013**, *117*, 3747-3761

**47)** Ganesh, P.; Jiang, D.; Kent, P.R.C. Accurate Static and Dynamic Properties of Liquid Electrolytes for Li-Ion Batteries from ab initio Molecular Dynamics. *J. Phys. Chem. B* **2011**, *115*, 3085-3090

**48)** Li, T.; Balbuena, P. Theoretical Studies of Lithium Perchlorate in Ethylene Carbonate, Propylene Carbonate, and Their Mixtures. *J. Electrochem. Soc.* **1999**, *146*, 3613-3622

**49)** Skarmoutsos, I.; Dellis, D.; Samios, J. Investigation of the local composition enhancement and related dynamics in supercritical $CO_2$-cosolvent mixtures: The case of ethanol in $CO_2$. *J. Chem. Phys.* **2007**, *126*, 224503

**50)** Errington, J. R.; Debenedetti, P. G. Relationship between structural order and the anomalies of liquid water. *Nature* **2001**, *409*, 318.

**51)** Kumar, P.; Buldyrev, S. V.; Stanley, H. E. A tetrahedral entropy for water. *Proc. Natl. Acad. Sci. USA* **2009**, *106*, 22130-22134.

**52)** Lazaridis, T.; Karplus, M. Orientational Correlations and entropy in liquid water. *J. Chem. Phys.* **1996**, *105*, 4294-4315.

**53)** Lazaridis, T.; Paulaitis, M. E. Simulation studies of the hydration entropy of simple, hydrophobic solutes. *J. Phys. Chem.* **1994**, *98*, 635-642.

**54)** Zielkiewicz, J. Two-particle Entropy and Structural Ordering in Liquid Water. *J. Phys. Chem. B* **2008**, *112*, 7810-7815.

**55)** Guardia, E.; Skarmoutsos, I.; Masia, M. Hydrogen Bonding and Related Properties in Liquid Water: A Car-Parrinello Molecular Dynamics Simulation Study. *J. Phys. Chem. B* **2014**, DOI: 10.1021/jp507196q

**56)** von Wald Cresce, A.; Xu, K. Preferential Solvation of Li$^+$ Directs Formation of Interphase on Graphitic Anode. *Electrochem. And Solid-State Lett.* **2011**, *14*, A154-A156.

**57)** Borodin, O. Molecular Modeling of Electrolytes. In *Electrolytes for Lithium and Lithium-Ion Batteries*; Jow, T.R.; Xu, K.; Borodin, O.; Ue, M., Eds.; Modern Aspects of Electrochemistry 58; Springer: New York, 2014, pp. 371-395.

**58)** Soetens, J.C.; Millot, C.; Maigret, B.; Bakó, I. Molecular Dynamics simulation and X-Ray diffraction studies of ethylene carbonate, propylene carbonate and dimethyl carbonate in liquid phase. *J. Mol. Liq.* **2001**, *92*, 201-216.

**59)** White, A.D.; Voth, G.A. Efficient and Minimal Method to Bias Molecular Simulations with Experimental Data. *J. Chem. Theory Comput.* **2014**, *10*, 3023-3030.

**60)** Soper, A.K. Computer simulation as a tool for the interpretation of total scattering data from glasses and liquids. *Molec. Simul.* **2012**, *38*, 1171-1185.